\documentclass[aps,rmp,reprint,amsmath,amssymb,graphicx,longbibliography]{revtex4-1}

\usepackage{bm}

\usepackage{graphicx,pstricks}
\usepackage{graphics}
\graphicspath{ {./images/} }
\usepackage{moreverb}
\usepackage{subfigure}
\usepackage{epsfig}
\usepackage{subfigure}
\usepackage{txfonts}
\usepackage{palatino}
\usepackage{caption}
\usepackage{amssymb}
\usepackage{amsmath}
\usepackage{qtree}
\usepackage{multirow}
\usepackage{comment}
\usepackage{geometry}
\usepackage{dsfont}
\usepackage[rightcaption]{sidecap}
\usepackage[thinc]{esdiff}
\usepackage{wrapfig}
\usepackage{graphicx}
\usepackage{dcolumn}
\usepackage{bm}
\usepackage{graphicx}
\usepackage{dcolumn}
\usepackage{bm}
\usepackage{mathrsfs}
\usepackage{tikz}
\usetikzlibrary{automata,arrows,positioning,calc}
\tikzset{node distance=4.5cm,   
         bend angle=15          
         }

\begin{document}

\title{Twenty-five years of random asset exchange modeling}

\author{Max Greenberg}
\email{Electronic address:
mg957@cornell.edu}
\affiliation{Department of Economics, University of Massachusetts Amherst, Amherst, MA 01002, USA}
\affiliation{Cornell Systems Engineering, Cornell University, Ithaca, NY 14850, USA}
\author{H. Oliver Gao} 
\email{Electronic address:
hg55@cornell.edu}
\affiliation{Cornell Systems Engineering, Cornell University, Ithaca, NY 14850, USA}

\date{\today{}}

\begin{abstract}
The last twenty-five years have seen the development of a significant literature within the subfield of econophysics which attempts to model economic inequality as an emergent property of stochastic interactions among ensembles of agents. In this article, the literature surrounding this approach to the study of wealth and income distributions, henceforth the "random asset exchange" literature following the terminology of Sinha (2003)\nocite{sinha_stochastic_2003}, is thoroughly reviewed for the first time. The foundational papers of Dr\u{a}gulescu and Yakovenko (2000), Chakraborti and Chakrabarti (2000), and Bouchaud and M\'{e}zard (2000) are discussed in detail, and principal canonical models within the random asset exchange literature are established. The most common variations upon these canonical models are enumerated, and significant papers within each kind of modification are introduced. The successes of such models, as well as the limitations of their underlying assumptions, are discussed, and it is argued that the literature should move in the direction of more explicit representations of economic structure and processes to acquire greater explanatory power.
\end{abstract}

\maketitle

\tableofcontents{}

\section{Introduction}

Over the last fifteen years, the question of economic inequality has become the epicenter of one of the most intense political debates in the United States. Awareness of the growing gap between rich and poor has been growing since the 2008 American bank bailouts and the 2010 \textit{Citizens United v. FEC} Supreme Court decision, but the inequality question was decisively pushed to the forefront of American politics in September 2011 with the beginning of the Occupy Wall Street protest movement, which introduced the dichotomy of "the 1\%" vs. "the 99\%" to public consciousness. Though the Occupy movement did not immediately produce anything by way of practical politics, it nonetheless laid the foundation for U.S. Senator Bernie Sanders' two campaigns for president, in which he used the language of Occupy to reframe economic inequality as the result of policy choices which could be rectified through a social-democratic "political revolution."

But the idea that economic inequality is a problem which needs to be addressed by way of policy is by no means an uncontroversial one. The majority of American adults who electorally favor the Republican Party do not believe that the current level of economic inequality in the United States is excessive \cite{pew_inequality_2020}. The legacy of "Reaganomics"---the economic policy pursued by the Federal government of the United States under the tenure of former President Ronald Reagan, which was characterized by cuts to tax rates and other concessions to proponents of "supply-side" economic theory---remains contentious. And within the Republican delegation to the United States House of Representatives, a proposal to eliminate the current bracketed income tax system and to replace it with a much higher nationwide flat sales tax, in order to dramatically lessen the tax burden on the wealthy, is gaining some traction \cite{dore_fair_2023}.

The controversy surrounding economic inequality is just as longstanding and just as intense within the realm of academic economics. On one hand, economists tend to be more skeptical than other social scientists of government intervention in the economy, as a great deal of emphasis is placed on the fact that, within the discipline's canonical models, free markets have no trouble arriving at socially optimal allocations of resources all on their own. On the other hand, French economist Thomas Piketty's 2013 magnum opus \textit{Capital in the 21st Century}, which proposes the imposition of a global, progressive tax on wealth in order to rein in what he views as the excessive amount of economic inequality in the world today, has become greatly influential in the popular-academic debates concerning the issue \cite{piketty_capital_2017}. Needless to say, the intense political disputes around the question of inequality show no sign of abating anytime soon.

\subsection{The universality of economic inequality}

What is indisputable, however, is that in nearly every single developed market economy, the degree of stratification between the rich and everyone else is not only staggering, but is also increasing. In the United States, the share of household wealth owned by the top 1\% of the population by net worth grew from 29.9\% in 1989 to 35.5\% in 2013; meanwhile, the share of wealth owned by the bottom 50\% of the population shrunk from 3.0\% to 1.1\% during the same period \cite{killewald_wealth_2017}. In Germany, individuals at the 90th percentile of net assets own 13 times as much wealth as the median individual and over a quarter of individuals have liabilities equal to or greater than their assets, resulting in a negative net worth \cite{grabka_persistently_2014}. While there are many countries where the degree of wealth inequality is not this extreme---the United States has one of the most unequal distributions of wealth in the world---the overall structure is strikingly similar in almost every country \cite{davies_level_2009}. In every market economy for which data exists, many possess very little wealth and a few possess much.

Great inequality also governs the distribution of incomes within market economies. As reported by Horowitz \textit{et al.} for the Pew Research Center, the share of aggregate income possessed by high-income households, defined as households with incomes greater than twice the national median, has grown from 29\% in 1970 to 48\% today\nocite{pew_inequality_2020}. In the same period, the share of aggregate income possessed by low-income households, defined as households with incomes less than two-thirds the national median, fell from 10\% to 9\% over the same period. In a similar vein, the incomes of those who are already in the top 5\% of the population in terms of earnings have consistently grown the faster than all other earners over the past 40 years. 

\subsection{Measuring inequality: the Gini coefficient}

One of the most popular metrics used to quantify the degree of inequality present in a given wealth or income distribution---or, indeed, any density distribution over a non-negative domain---is the Gini coefficient, named for Italian statistician Corrado Gini. The Gini coefficient of a distribution \(f(x)\) is defined by reference to the Lorenz curve, itself defined as the function: 
\begin{equation}
    L\left (F(x) \right ) = \frac{1}{\mu} \int_{0}^{x} s \cdot p(s) ds
\end{equation}
where \(F(x) = \int_{0}^{x} f(s) ds\) is the cumulative density distribution of \(f(x)\) and \(\mu = \int_{0}^{\infty} s \cdot f(s) ds\) is the mean of \(f(x)\) \cite{gastwirth_general_1971}. Intuitively, this integral represents the share of some asset---say, income---held by the bottom \(100x\)\% of the population, normalized by the mean of the distribution. The Gini coefficient is then given by twice the difference between the area under Lorenz curve of a perfectly egalitarian distribution---a straight line with a slope of 1---and the Lorenz curve of the distribution in question \cite{dorfman_formula_1979}. Thus, the canonical formula used to calculate the Gini coefficient is:
\begin{equation}
    G = 1 - 2 \int_{0}^{1} L(x) dx
\end{equation}
Thus, the Gini coefficient can take on any value between 0---perfect equality---and 1---perfect inequality. To extend this statistic to describe dispersion within a finite population \(\{x_{i}\}_{i=1}^{N}\), however, it is more convenient to leverage the alternative (but equivalent) definition of the Gini coefficient, given by half the relative mean absolute difference of a distribution: 
\begin{equation}
    G = \frac{1}{2} \frac{\left ( \sum_{i=1}^{N} \sum_{j=1}^{N} |x_{i} - x_{j}| \right ) / N^{2}}{\mu}
\end{equation}
If the population \(\{x_{i}\}_{i=1}^{N}\) is sorted such that \(x_{j} > x_{k}\) if and only if \(j > k\), the Gini coefficient may be calculated using the computationally much faster formula:
\begin{equation}
    G = \frac{N + 1}{N} - \frac{2}{\mu N^{2}} \sum_{i=1}^{N} (N - i + 1) x_{i} 
\end{equation}
as demonstrated by Allison (1978).\nocite{allison_measures_1978} Note that the Gini coefficient over a discrete population does not perfectly correspond to its continuous counterpart, however, as the former has an upper bound of \(1 - 1/N\) \cite{allison_reply_1979}.

The Gini coefficent is not a perfect measure of inequality. It has been criticized on the basis that distributions with very different levels of concentration in the right tail can produce identical indices; there is therefore significant information lost when using it to represent an entire distribution with a single scalar value \cite{osberg_limitations_2017}. Nonetheless, the Gini coefficient serves as a useful and widely-used benchmark for summarizing the degree of dispersion present in a given wealth or income distribution.

\begin{table*}[t]
    \centering
\begin{tabular}{ |p{3.5cm}||p{4.5cm}|p{4.5cm}| }
 \hline
 \multicolumn{3}{|c|}{\textbf{Gini Coefficients of Wealth and Income for Ten Countries}} \\
 \hline
 \textit{Country} & \textit{Gini Coefficient of Wealth} & \textit{Gini Coefficient of Income} \\
 \hline
United States   & 0.801 & 0.401 \\
France          & 0.730 & 0.311 \\
United Kingdom  & 0.697 & 0.396 \\
India           & 0.669 & 0.344* \\
Germany         & 0.667 & 0.289 \\
Netherlands     & 0.650 & 0.298* \\
Australia       & 0.622 & 0.331* \\
Italy           & 0.609 & 0.353 \\
Spain           & 0.570 & 0.343 \\
China           & 0.550 & 0.420* \\
\hline
\end{tabular}
    \caption[Gini coefficients of wealth and income inequality for ten countries]{Gini coefficients of wealth and income inequality for ten countries, all major world economies, based on data from the year 2000. Data for Gini coefficients of wealth are taken from Davies \textit{et al.} (2009), while data for Gini coefficients of income are taken from the FRED database hosted by the Federal Reserve Bank of St. Louis. Gini coefficients of income for India, the Netherlands, and Australia are from 2004. Gini coefficient of income for China is from 2002.}
    \label{tab:gini_table}
\end{table*}

Underscoring the universality of steep economic inequality in both wealth and income distributions, Table \ref{tab:gini_table} displays the Gini coefficients for the wealth and income distributions of ten countries. We observe that wealth distributions are almost always "more unequal" than income distributions: Gini coefficients for wealth distributions tend to range between \(0.5\) and \(0.8\), while Gini coefficients for income distributions tend to range from \(0.25\) to \(0.45\). Furthermore, there is no obvious correlation between the Gini coefficients for wealth distributions and for income distributions: some countries, such as China, have coefficients relatively close in value, while other countries, such as France, have Gini coefficients for wealth over twice as high as the corresponding value for income.

\subsection{Pareto, Gibrat, and the econophysicists}

Regardless of whether one personally believes governments should play a role in redistributing wealth from the rich to the poor, the universality of the phenomenon of extreme inequality should raise eyebrows. Different countries have dramatically different approaches to welfare programs, taxation, and all other sorts of policy. Yet the distributions of wealth and income which emerge in these countries are remarkably similar in form. It follows that there must be some shared set of characteristics that account for this common structure of wealth distribution. This line of questioning points us to an often overlooked and still poorly understood aspect of economic inequality: its origin.

The nature and origin of the distribution of wealth and income in market economies has been an open problem in economics for more than a century. In 1897, the Italian civil engineer-turned-economist Vilfredo Pareto attempted to provide an answer after noticing a striking pattern in data for land-ownership rates in Italy. Specifically, Pareto posited that income in every society was distributed according to a decreasing power law; namely:
\begin{equation}
    p(x) \propto x^{-1-\alpha}
\end{equation}
where \(p(x)\) represents the probability density function of income and \(\alpha\) represents the "Pareto index," with smaller values producing fatter tails and thus representing more unequal distributions. This observation has come to be known as the "weak Pareto law," with its strong counterpart including the additional claim that the Pareto index possesses a value in the range \(1.5 \pm 0.5\) \cite{pareto_cours_1897}. But not long thereafter it became apparent that this law did not actually well characterize the entire income distribution. Instead, when low- and middle-income strata were taken into account, the data seemed to be much better fit by a right-skewed lognormal distribution:
\begin{equation}
    p(x) = \frac{1}{x \sigma \sqrt{2 \pi}} \exp \left ( - \frac{(\ln(x) - \mu)^{2}}{2 \sigma^{2}} \right )
\end{equation}
This fact was first noticed by Robert Gibrat \cite{gibrat_les_1931}. It is now well established that, in fact, both Pareto and Gibrat were correct: a lognormal-like distribution tends to characterize the bulk of incomes, while the Pareto distribution tends to characterize the highest 2-3\% of incomes \cite{montroll_1f_1982}.

Since these discoveries, mainstream economic theory has, broadly speaking, shied away from further attempts to impose a universal form to these distributions or to explain the processes responsible for their emergence. There are both normative and methodological reasons for this gap in the economics literature. The normative aversion, as voiced by Piketty, manifests as skepticism that a universal law governing the right tails of income distributions exists at all \cite{piketty_capital_2017}. The methodological aversion, on the other hand, stems from the fact that most macroeconomic models make use of single, representative agents, which are ill-suited for describing heterogeneity within a population. Meanwhile, more sophisticated tools capable of addressing such questions, such as the Heterogeneous Agent New Keynesian (HANK) class of models, are still relatively new to the scene \cite{achdou_income_2022}. Nonetheless, this aversion has meant that there has remained comparatively little in the way of literature concerning one of the most crucial questions in economics today. This gap drew the attention of physicists interested in applying methods developed for the study of the natural sciences to questions in the social sciences in the late 1990s.

The aim of these "econophysicists" was to capture the characteristic features of empirical wealth and income distributions, as made known by extensive statistical analyses. There is now substantial evidence that the bulk of the income distribution in all capitalist countries follows an exponential distribution \cite{tao_exponential_2019}. The right tail of the income distribution follows the aforementioned Pareto law and the left tail follows Gibrat's law. The exponential bulk and the log-normal left tail are sometimes unified in the form of the closely related Gamma distribution:
\begin{equation}
    p(x) = \frac{\beta^{\alpha}}{\Gamma(\alpha)} x^{\alpha - 1} e^{-\beta x}
\end{equation}
where \(\alpha\) is called the "shape parameter" and \(\beta\) the "rate parameter." However, there remains insufficient data to conclude whether the Gamma or log-normal distribution provides the better empirical fit \cite{chakrabarti_econophysics_2013}.

Wealth distributions are unfortunately much less well understood due to a dearth of publicly available data. Rough estimates of wealth distributions in pre-capitalist societies, such as in the New Kingdom of Egypt and medieval Hungary, provide some evidence that such societies exhibited power-law distributions of wealth, but these results are far from conclusive \cite{abul-magd_wealth_2002, hegyi_wealth_2007}. Dr\u{a}gulescu and Yakovenko (2001b) used inheritance tax data to study the wealth distribution in the modern United Kingdom, which was found to have a similar structure to the UK's income distribution.\nocite{dragulescu_exponential_2001} Further supporting this conclusion, Sinha (2006), among others, found evidence that the very wealthiest stratum of society, as measured by published "rich lists," follows a power law distribution as well.\nocite{sinha_evidence_2006} These features appear to emerge even in artificial economies, with Fuchs \textit{et al.} (2014) having observed an exponential bulk and power-law tail even in the wealth distribution across players of a massively multiplayer online game with inbuilt systems of production and trade.\nocite{fuchs_behavioral_2014} Thus, early exchange models in the econophysics literature sought to generate distributions exhibiting both the exponential bulk and power-law tail observed in data by means of symmetric binary interactions.

The first paper in this lineage was Ispolatov \textit{et al.} (1998), and shortly thereafter two papers which would ultimately become the cornerstones of the random asset exchange modeling literature---Dr\u{a}gulescu and Yakovenko (2000) and Bouchaud and M\'{e}zard (2000)---emerged.\nocite{ispolatov_wealth_1998, dragulescu_statistical_2000, bouchaud_wealth_2000} As it turned out, however, the econophysicists were not the first to approach the question of inequality in this way. The sociologist John Angle had actually published a series of papers containing a model extremely similar to Ispolatov \textit{et al.}'s more than a decade earlier, though the literature had no knowledge of this fact until it was pointed out by Lux (2005) \cite{angle_surplus_1986, angle_inequality_1992, angle_deriving_1993}.\nocite{lux_emergent_2005} Likewise, it was noticed by Patriarca \textit{et al.} (2005) that Dr\u{a}gulescu and Yakovenko's model was anticipated by a series of papers by Eleonora Bennati, which had been published in ill-known Italian economics journals in the 1980s and which had not been translated into English \cite{bennati_metodo_1988, bennati_metodo_1993}.\nocite{patriarca_kinetic_2005}

Nonetheless, in the twenty-five years since Ispolatov \textit{et al.}'s initial paper, a sizeable literature on this subject has emerged, with countless variations of the aforementioned models having been proposed and investigated. The literature has also become much more diverse in that time: though this subject was initially solely the domain of a subset of physicists interested in exploring economic questions, they have since been joined by researchers with backgrounds in mathematics, economics, systems science, and more.

This article provides, for the first time, a comprehensive and thoroughgoing review of this literature, which, following the terminology of Sinha (2003), will be referred to as the "random asset exchange" literature. While many excellent partial reviews do already exist (see Chatterjee and Chakrabarti
(2007), Yakovenko and Rosser (2009), Patriarca
et al. (2010), and Patriarca and Chakraborti
(2013), just to name a few\nocite{chatterjee_kinetic_2007, yakovenko_colloquium_2009, patriarca_microscopic_2017, patriarca_kinetic_2013}), all either have become dated or have focused on only a delimited part of the literature. This review is the first, to our knowledge, that not only discusses all significant econophysical models of income inequality, but fully enumerates all the most common variations upon the literature's canonical models as well. 

\section{The Taxonomy of Random Asset Exchange Models}

Most random asset exchange models tend to fall into one of two classes. The first of these is conventionally called the "kinetic wealth exchange" (KWE) class of model, which was popularized by Dr\u{a}gulescu and Yakovenko (2000). Named such because of their similarity to thermodynamic models from the kinetic theory of gases, KWE models are typically---though not always---characterized by the following properties:
\begin{enumerate}
    \item Pairwise exchange between agents is the primary system state transition function;
    \item Total money present in the system is conserved; and
    \item Total money present between all pairs of agents engaged in exchange is conserved.
\end{enumerate}
These features are analogous to the role of particle collisions, conservation of energy, and conservation of momentum in the kinetic theory of gases, respectively. 

The second prominent class of model, inspired by models of directed polymers rather than ideal gases, is the Bouchaud-M\'{e}zard (BM) type model, first introduced by Bouchaud and M\'{e}zard (2000).\nocite{bouchaud_wealth_2000} In contrast to KWE-style models, BM-style models tend to be characterized instead by fixed wealth flow rates between all "adjacent" pairs of agents, as defined by an implicit adjacency network, being the main mechanism of system evolution. Furthermore, each agent's wealth is subject to endogenous stochastic variation, leading to systemic non-conservation of wealth.

While other formulations exist, models belonging to one of these two classes represent the great majority of the literature. In this section, we review the most significant variations of both classes---as well as a handful of minor but nonetheless significant alternative model classes.

\subsection{Kinetic wealth exchange}

While their work was anticipated by Angle (1986), Bennati (1988, 1993), and Ispolatov \textit{et al.} (1998), it is Dr\u{a}gulescu and Yakovenko (2000) who are credited with first formalizing and thoroughly studying the KWE model.\nocite{angle_surplus_1986, bennati_metodo_1988, bennati_metodo_1993, ispolatov_wealth_1998, dragulescu_statistical_2000} In their initial formulation, a system of \(N \gg 0\) agents with \(M \gg N\) units of wealth between them is posited. Agents then engage in random pairwise exchanges, with a winner and loser being randomly selected in each pair and a transfer of wealth occurring, following some exchange rule:
\begin{equation}
    \begin{bmatrix} w_{i} \\ w_{j} \end{bmatrix} \rightarrow \begin{bmatrix} w_{i} + \Delta w \\ w_{j} - \Delta w \end{bmatrix}
\end{equation}
with \(\Delta w > 0\) if agent \(i\) is the winner of the exchange, and \(\Delta w < 0\) if instead \(j\) is victorious. Since KWE models almost always feature exclusively linear, pairwise exchanges, it is often convenient to represent the model's exchange rule as a \(2 \times 2\) matrix \(M\), such that:
\begin{equation}
    \begin{split}
        \begin{bmatrix}w_{i}(t+1) \\ w_{j}(t+1) \end{bmatrix} &= M\begin{bmatrix}w_{i}(t) \\ w_{j}(t) \end{bmatrix}
    \end{split}
\end{equation}

Dr\u{a}gulescu and Yakovenko showed that, so long as \(\Delta w\) is chosen so that the exchange process was time-reversal symmetric, then the distribution of money among agents converges to the entropy-maximizing exponential distribution:
\begin{equation}
    p(w) = \frac{1}{T}\exp \left ( -\frac{w}{T} \right )
\end{equation}
where \(T = \langle w \rangle\) represents the average wealth held by agents---analogous to temperature in the equivalent thermodynamic system \cite{dragulescu_statistical_2000}. This result proves to be extremely robust, not varying with one's choice of time-reversal symmetric exchange rule or underlying adjacency network \cite{lanchier_rigorous_2017}.

The differences between Dr\u{a}gulescu and Yakovenko's model and those of Angle, Ispolatov \textit{et al.}, and Bennati are subtle. In both Angle's initial model (the "one-parameter inequality process," or OPIP) and Ispolatov \textit{et al.}'s "multiplicative-random" exchange model, \(\Delta w = \varepsilon w_{loser}\), such that exchanges are of the form:
\begin{equation}
    \begin{bmatrix} w_{i} \\ w_{j} \end{bmatrix} \rightarrow \begin{bmatrix} w_{i} + \varepsilon w_{j} \\ (1 - \varepsilon) w_{j} \end{bmatrix}
\end{equation}
if agent \(i\) wins the exchange. The sole difference between these two formulations is that Angle (1986) draws \(\varepsilon\) from a uniform distribution before each exchange, whereas Ispolatov \textit{et al} (1998) define \(\varepsilon\) as a fixed, global parameter. In contrast to the exchange rules investigated by Dr\u{a}gulescu and Yakovenko (2000), both of these models break time symmetry and produce identical distributions which are very well-approximated by, but not exactly given by, Gamma distributions \cite{angle_deriving_1993}.

In both Ispolatov \textit{et al.}'s additive-random exchange model and Bennati's model, on the other hand, agents exchange constant, quantized amounts of wealth, equivalent under rescaling to \(\Delta w = 1\). In Ispolatov \textit{et al.} (1998), agents with 0 wealth are removed from the system entirely, causing all the wealth in the system to eventually be accumulated by a single agent (a phenomenon termed "condensation"). In Bennati (1988), however, agents with 0 wealth are permitted to win, but not to lose, exchanges, identical to the provision in the constant exchange rule discussed by Dr\u{a}gulescu and Yakovenko. For that reason, the KWE model with time reversal-symmetric exchange rule is sometimes referred to as the Bennati-Dr\u{a}gulescu-Yakovenko (BDY) model of wealth exchange \cite{yakovenko_colloquium_2009}.

An extension of Dr\u{a}gulescu and Yakovenko's initial model proposed contemporaneously with its initial publication was investigated by Chakraborti and Chakrabarti (2000), which introduced a "saving propensity" parameter \(\lambda\).\nocite{chakraborti_statistical_2000} Called the CC model (or, more rarely, the "saved wealth" model), its system dynamics are characterized by the fact that, for \(\lambda \in [0, 1)\), every agent engages in multiplicative exchange with only a fraction \(1 - \lambda\) of their total wealth. The exchange rule in such models can thus be defined by:
\begin{equation}
    M = \begin{bmatrix} \lambda + \varepsilon (1 - \lambda) & \varepsilon (1 - \lambda) \\ (1 - \varepsilon) (1 - \lambda) & \lambda + (1 - \varepsilon) (1 - \lambda) \end{bmatrix}
\end{equation}
where \(\varepsilon\) is drawn from a uniform distribution on \([0,1]\) at every exchange. 

Curiously, this slight modification dramatically changes the equilibrium distribution of money among agents within the system as the mode of the distribution (the "most likely agent wealth") becomes non-zero, approaching \(T = \frac{M}{N}\) (an egalitarian distribution) as \(\lambda\) approaches 1. Gupta (2006) observes that this departure from the entropy-maximizing distribution is a consequence of the fact that the introduction of the saving propensity parameter \(\lambda\) results in the system transition matrix becoming non-singular \cite{gupta_money_2006}. Patriarca \textit{et al.} (2004a, b) demonstrate that the resultant distribution is extremely well fit by a scaled Gamma distribution:
\begin{equation}
    \begin{split}
        p(w) = \frac{1}{\Gamma(n)} \left ( n \cdot \frac{w}{T} \right )^{n-1} \exp \left (- n \cdot \frac{w}{T} \right )
    \end{split}
\end{equation}
where \(n = 1 + 3\lambda/(1 - \lambda)\).\nocite{patriarca_statistical_2004, patriarca_gibbs_2004} The fit is not exact, however, as the distributions differ in their fourth moments \cite{lallouache_wealth_2010}.

The CC model is extremely influential in the random asset exchange literature, and it itself has two major variations which must be mentioned. The first, introduced by Chatterjee \textit{et al.} (2003), defines the saving propensity parameter to be heterogenously distributed throughout the population; instead of having identical saving propensities, each agent \(i\) has their own individual saving propensity \(\lambda_{i} \in [0, 1)\) drawn from the uniform distribution during model initialization.\nocite{chatterjee_money_2003} The exchange rule of this model, called the CCM model, is thus:
\begin{equation}
    M = \begin{bmatrix} 
    \lambda_{i} + \varepsilon (1 - \lambda_{i}) & \varepsilon (1 - \lambda_{j})\\
    (1 - \varepsilon)(1 - \lambda_{i}) & \lambda_{j} + (1 - \varepsilon) (1 - \lambda_{j})
    \end{bmatrix}
\end{equation}
The steady-state distribution then exhibits a Gamma-like bulk, as in the CC model, as well as a right tail well fit by a power law with Pareto parameter \(\alpha = 1\); this power law is robust for any distribution of saving propensity of the form \(\rho(\lambda) \approx |\lambda_{0} - \lambda|^{\alpha}\), or for uniform distributions within a restricted range \(\lambda_{i} \in [a, b] \subset [0, 1)\) \cite{chatterjee_pareto_2004}.

One well-known (and, arguably, unrealistic) aspect of the CCM model is that average agent wealth is highly correlated with their saving parameter, such that the agents who save nearly all of their money in every transaction invariably become the wealthiest. This remains the case even if a significant bias in favor of poorer agents is introduced, because thrifty agents in the CCM model always stand to gain much more than they lose from every transaction \cite{nener_optimal_2021}. This aspect of the model also explains the surprising appearance of the Pareto tail, which is actually somewhat illusory: the right tail of the equilibrium distribution of the CCM model is constituted by the overlapping exponential tails of the exponential distributions corresponding to the subpopulations with the highest saving parameters \cite{patriarca_kinetic_2005}.

Another significant drawback of the CCM model is that, while the right tails of the steady state distributions change from approximately Pareto with index 1 to exponential as the distribution of \(\lambda_{i}\) narrows, the empirical value of \(\alpha \approx 1.5\) is never reached \cite{repetowicz_dynamics_2005}. However, there are a number of ways to modify the CCM model and recover such a regime: Repetowicz \textit{et al.} (2006), for instance, note that introducing modified wealth parameters with memory---\(\hat{w}_{i}(t) = w_{i}(t) + \gamma w_{i}(q)\), with \(\gamma \in (0, 1)\) and \(q < t\)---before each transaction and applying the CC exchange rule thereto does permit Pareto tails with indices \(\alpha > 1\) to be obtained.\nocite{repetowicz_agent_2006} Likewise, Bisi (2017) demonstrates that replacing the saving propensity parameter with a bounded, global function of an agent's wealth \(\gamma(w_{i})\) also permits superunitary Pareto indices.\nocite{bisi_kinetic_2017}

\begin{figure*}
    \centering
    \includegraphics[width=\textwidth,height=\textheight]{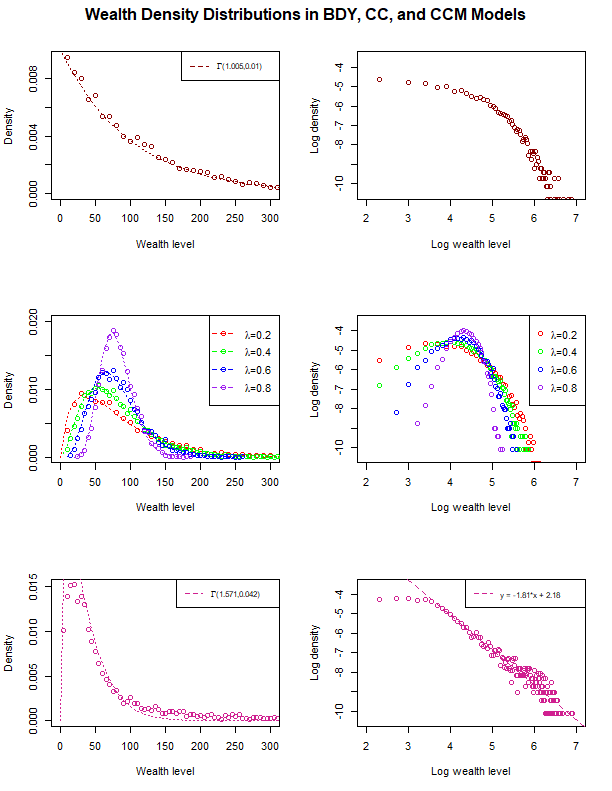}
    \caption[Stationary distributions produced by BDY, CC, and CCM models]{Stationary distributions produced by the BDY (left), CC (middle), and CCM (right) models, with best-fit gamma curves shown. All simulations were performed with the parameterization \(N = 5000\) and \(\langle w \rangle = 100\) over \(10^{5}\) iterations.}
    \label{fig:bdy-cc-ccm_plot}
\end{figure*}

The second variation, introduced by Cordier \textit{et al.} (2005) and often called the CPT model, investigates the CC model with an additional stochastic growth term:
\begin{equation}
    \begin{split}
        M
        &= 
        \begin{bmatrix}
        (1 - \lambda) + \eta_{i} & \lambda \\
        \lambda & (1 - \lambda) + \eta_{j}
        \end{bmatrix}
    \end{split}
\end{equation}
Where \(\eta_{i}\) and \(\eta_{j}\) are independent and identically distributed variables with mean 0 and variance \(\sigma^{2}\).\nocite{cordier_kinetic_2005} As in the BDY, CC, and CCM models, debts are not permitted, so a transaction only takes place so long as neither agent is reduced to a negative level of wealth. Because \(\eta_{i}\) and \(\eta_{j}\) are uncorrelated, total wealth is now only preserved in the mean. The CPT model has an inverse-Gamma equilibrium, with shape parameter \(\alpha = 1 + \frac{2\lambda}{\sigma^{2}}\) and scale parameter \(\beta = \alpha - 1\):
\begin{equation}
    \begin{split}
        p(w) &= \frac{\beta^{\alpha}}{\Gamma(\alpha)} \cdot w^{-1 - \alpha} \cdot \exp \left ( - \frac{\beta}{w} \right )
    \end{split}
\end{equation}
In this case, the shape parameter \(\alpha\) may be interpreted as the "Pareto index" of the approximately power-law right tail.

The CPT model, like the CCM model, is quite flexible and has been studied in a variety of other contexts. D\"{u}ring and Toscani (2008) employ a CPT model with quenched saving propensities to study international transactions, representing countries as subpopulations with different saving propensities.\nocite{during_international_2008} Bisi and Spiga (2010) consider a variation on the CPT wherein the amount of wealth an agent receives from his trading partner is also subject to stochastic fluctuations.\nocite{bisi_boltzmann-type_2010} More recently, Zhou \textit{et al.} (2021) investigate the effect of introducing a non-Maxwellian (i.e. wealth-varying) collision kernel in the CPT model.\nocite{zhou_study_2021}

\subsubsection{Theft, fraud, and yard sales}

Following the terminology of Hayes (2002), binary exchange models in which the transfer amount is proportional to the wealth of the loser are commonly referred to as "theft and fraud" (TF) models, while those in which the transfer amount is proportional to the wealth of the poorer agent are referred to as "yard sale" (YS) models. That is, the YS model posits an exchanged quantity \(\Delta w\) of the form:
\begin{equation}
    \begin{split}
        \Delta w &\propto \min \{ w_{i}, w_{j} \}
    \end{split}
\end{equation}
The advantage of the YS model is that, from a strategic perspective, agents are not disincentivized from engaging in trade, as the expected value of an exchange is always 0. This is in contrast to the TF model, which is so-named precisely because the expected value of an exchange is always negative for the richer agent. If agents were allowed to choose whether or not to engage in a given exchange, a TF economy would immediately freeze as soon as a wealth differential appeared. The principal drawback of the YS model, however, is that it is now well-known that the unmodified YS model always exhibits condensation, though a non-degenerate equilibrium can be recovered if the probability of winning a given exchange is biased in favor of the less wealthy agent, if a mechanism for redistributing wealth from richer agents to poorer ones is introduced, or if extremal dynamics are coupled to the system \cite{boghosian_h_2015, sinha_stochastic_2003, cardoso_wealth_2021, bagatella-flores_wealth_2015}.

Moukarzel \textit{et al.} (2007) demonstrated that, in the case of the YS model where the proportion of the poorer agent's wealth at stake in each transaction is a fixed constant \(f\), a sufficient bias of the probability \(p\) towards the poorer agent alone was sufficient to avoid condensation.\nocite{moukarzel_wealth_2007} In particular, the critical probability \(p^{*}\) above which the system does not condense was found to be:
\begin{equation}
    p^{*} = \frac{\log\left ( \frac{1}{1 - f} \right )}{\log \left ( \frac{1 + f}{1 - f}\right )}
\end{equation}
Based on this result, Bustos-Guajardo and Moukarzel (2012) studied an extension of the YS model on an adjacency network, such that exchanges may only take place between adjacent agents. They found that the value of the critical probability remains the same regardless of the choice of network.\nocite{bustos-guajardo_yard-sale_2012} In fact, most system dynamics in the stable phase of the system are independent of the choice of network. However, certain dynamical aspects of the system (such as time required for the system to fully condense) do differ from the fully-connected case in the unstable (i.e. condensing) region. This is not entirely surprising, seeing as the number of agents to whom the wealth will condense is directly determined by the underlying network; instead of one agent accumulating all the money, the distribution condenses to a set of "locally rich agents," sometimes termed the "oligarchy."

Redistribution in the YS model was examined by Boghosian (2014a), in which a mechanism by which, at each time step, \(\chi\) percent of each agent's wealth was confiscated and subsequently redistributed uniformly among the population.\nocite{boghosian_kinetics_2014} Introducing this mechanism not only prevented condensation, but also produced a gamma-like steady state distribution with a Pareto-like tail. The dynamics of this mechanism were studied in more detail by Boghosian \textit{et al.} (2017) and Devitt-Lee \textit{et al.} (2018), in which it was combined with an bias in exchanges in favor of the wealthy, called the "wealth-attained advantage."\nocite{boghosian_oligarchy_2017, devitt-lee_nonstandard_2018} In this variation, termed the "extended yard sale" (EYS) model, the wealthier agent wins a given exchange with probability \(p = rT(w_{i} - w_{j})\), where \(T\) is the average wealth of the system and \(w_{i}\) is the wealth of the richer agent. The wealth-attained advantage formally acts as a net tax on the non-oligarchy, while the redistribution acts as a net tax on the oligarchy; once the poor-to-rich flux of the redistributive mechanism was eclipsed by the rich-to-poor flux of the wealthy agents' advantage, the system moves from a subcritical to a supercritical state and the inequality of the resulting wealth distribution, as measured by the Gini coefficient, begins increasing rapidly. This acts as a second-order phase transition within the system. One additional variation of the EYS model, the "affine wealth" (AW) model, was introduced by Li \textit{et al.} (2019).\nocite{li_affine_2019} The AW model permits negative wealth by defining a debt limit \(\Delta\), adding \(\Delta\) to the wealths of both agents before each exchange, and subtracting \(\Delta\) once the exchange is complete. The AW model provides a remarkably good fit to the U.S. wealth distribution, as reported by the U.S. Survey of Consumer Finances.

\begin{figure*}
    \centering
    \includegraphics[width=\textwidth]{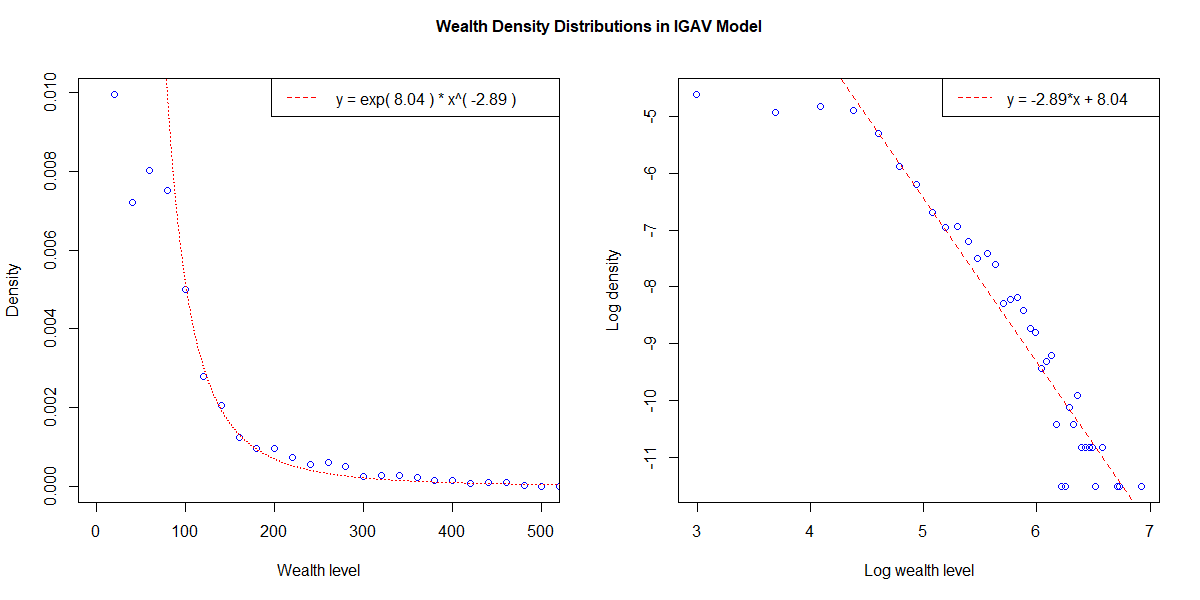}
    \caption[Stationary distribution produced by IGAV model]{Stationary distribution produced by the IGAV model with quenched savings propensities and a bias function per Eq. (12). Simulation was performed with the parameterization \(N = 5000\) and \(\langle w \rangle = 100\) over \(10^{6}\) iterations.}
    \label{fig:igav_plot}
\end{figure*}

It is also worth mentioning the variation on the YS model first formulated by Iglesias \textit{et al.} (2004).\nocite{iglesias_correlation_2004} This model, sometimes referred to as the IGAV model, sees agents with wealths \(w_{i}\) and \(w_{j}\) and saving parameters \(\lambda_{i}\) and \(\lambda_{j}\) exchange quantities \(\Delta w_{ij} = \min \left \{ (1 - \lambda_{i})w_{i}, (1 - \lambda_{j})w_{j} \right \}\), where the bias in favor of the poorer agent is defined as per Scafetta \textit{et al.} (2002)\nocite{scafetta_paretos_2002}:
\begin{equation}
    p = \frac{1}{2} + f \cdot \frac{w_{1} - w_{2}}{w_{1} + w_{2}}
\end{equation}
The asymmetry flux index \(f \in [0, 1/2]\) essentially defines the degree of systemic social protection offered to the poor. A number of similar models which modify Iglesias \textit{et al.}'s exchange rule were studied by Caon \textit{et al.} (2007).\nocite{caon_unfair_2007} Recently, Neñer and Laguna (2021a) showed that, in the IGAV model, the richest agents are not necessarily the thriftiest.\nocite{nener_optimal_2021} Instead, the saving propensity \(\lambda^{*}_{i}\) that maximizes equilibrium average wealth lies in the interval \((0, 1)\), increasing with \(f\).

Heinsalu and Patriarca (2014) introduce a variation of the BDY model meant to more explicitly model the dynamics of barter economies.\nocite{heinsalu_kinetic_2014} While Dr\u{a}gulescu and Yakovenko examined, among others, the TF exchange rule:
\begin{equation}
    M = \begin{bmatrix} \varepsilon & \varepsilon \\ 1 - \varepsilon & 1 - \varepsilon \end{bmatrix}
\end{equation}
where \(\varepsilon\) is a uniform random variable with mean \(0.5\), Heinsalu and Patriarca consider the rule:
\begin{equation}
    M = \begin{bmatrix} 1 - \varepsilon_{i} & \varepsilon_{j} \\ \varepsilon_{i} & 1 - \varepsilon_{j} \end{bmatrix}
\end{equation}
where \(\varepsilon_{i}\) and \(\varepsilon_{j}\) are i.i.d. uniform random variables with mean \(0.5\). This modification, called the "immediate exchange" (IE) model, changes the system from a pure TF one, where wealth flows unidirectionally and, on average, from richer to poorer agents, to one in which wealth flows bidirectionally. In the general IE model, transactions have some probability \(\mu\) of occurring unidirectionally in the manner of Angle (1986). In the pure IE model with \(\mu = 0\) has a steady state distribution \(p(w)\) which is an exact Gamma distribution with a shape parameter of 2, meaning \(\lim_{x \rightarrow 0} p(x) = 0\), \(p(x)\) has a non-zero mode, and the right tail is well-approximated by a Pareto distribution with \(\alpha = 1\) \cite{katriel_immediate_2014}.

\subsection{Bouchaud-M\'{e}zard models}

Unlike the KWE model, the BM model does not make use of agents pairing up and engaging in binary transactions with a winner and a loser; rather, the rate of exchange between agents are defined by a fixed adjacency matrix \(\mathbf{J}\), each entry of which \(J_{ij}\) represents the "cash flow rate" from agent \(j\) to agent\(i\). In Bouchaud and M\'{e}zard's original paper, each agent in the population of size \(N\) has two sources of income---stochastic returns from investments and sales of a product to other agents---and one source of expenses---purchases of products from other agents. Thus, the income of agent \(i\) is given by:
\begin{equation}
    \diff{w_{i}}{t} = \eta_{i}(t) w_{i}(t) + \sum_{j \neq i} J_{ij}w_{j}(t) - \sum_{j \neq i} J_{ji} w_{i}(t)
\end{equation}
where \(\eta_{i}\) is a Gaussian random variable with variance \(2\sigma^{2}\) \cite{bouchaud_wealth_2000}. Notably, the BM model has no restriction on total wealth being conserved. The simplest case, in which all rates of exchange are equalized such that \(J_{ij} = \frac{J}{N}\), lends itself well to a mean-field approximation, which produces an inverse-Gamma equilibrium distribution with shape parameter \(\alpha = 1 + \frac{J}{\sigma^{2}}\) and scale parameter \(\beta = \alpha - 1\)---strikingly similar in form to the equilibrium distribution of the CPT model. However, the mean-field approximation is time-limited; for any finite number of agents, the BM model on a complete graph will eventually exhibit wealth condensation and the probability that a given agent will have wealth less than any finite fraction of total wealth grows to 1 \cite{medo_breakdown_2009}.

Further investigation into this class of model demonstrated that the resulting distribution is also sensitive to the nature of the underlying network defining the non-zero entries of the transaction matrix \(\mathbf{J}\). Souma \textit{et al.} (2001) demonstrated through simulation that defining \(\mathbf{J}\) on a small-world network---where each agent neighbors only \(0.1\%\) of the population---leads to distributions which are best fit by a combination of log-normal and power-law distributions.\nocite{souma_small-world_2001} Garlaschelli and Loffredo (2004, 2008) likewise showed that it is possible to retrieve a realistic mixed log-normal-power law distribution by simulating the model on a simple heterogeneous network with a small number of "hub" agents, and that the BM model on a homogeneous network is able to reproduce either a log-normal or a power law distribution---but not both---depending on the average number of adjacencies per agent.\nocite{garlaschelli_wealth_2004, garlaschelli_effects_2008} Ma \textit{et al.} (2013) simulated the BM model on a partially connected network and found the generalized inverse Gamma (GIGa) distribution provided the best fit to the steady state.\nocite{ma_distribution_2013}

Though the original BM model is a continuous-time model, a number of authors have studied similar models in discrete time as well. Di Matteo \textit{et al.} (2003), for example, considers the variation\nocite{di_matteo_exchanges_2003}:
\begin{equation}
    \begin{split}
        \Delta w_{i}(t) &= A_{i}(t) + B_{i}(t)w_{t} \\ &+ \sum_{j \neq i} Q_{j \rightarrow i}(t) w_{j}(t) - \sum_{j \neq i} Q_{i \rightarrow j}(t) w_{i}(t)
    \end{split}
\end{equation}
For the purposes of their analysis, additive noise \(A_{i}(t)\) is assumed to be Gaussian with mean zero, and multiplicative noise \(B_{i}(t) = 0\). Additionally, each agent \(i\) is assumed to split a fixed share \(q_{0}\) of their wealth evenly with all of their neighbors \(j \in \mathcal{I}_{i}\), where \(|\mathcal{i}| = z_{i}\). Thus, \(Q_{i \rightarrow j}(t) = \frac{q_{0}}{z_{i}}\) if \(j \in \mathcal{I}_{i}\) and 0 otherwise. Their restricted system dynamics thus become:
\begin{equation}
    \begin{split}
        w_{i}(t+1) - w_{i}(t) &= A_{i}(t) - q_{0} w_{i}(t) + \sum_{j \in \mathcal{I}_{i}} \frac{q_{0}}{z_{j}} w_{j}(t)
    \end{split}
\end{equation}
Simulating this time series produces results dependent on the choice of adjacency network, the most notable being that scale-free networks produce power-law distributions. In this case, the equilibrium wealth level of a given node is nearly perfectly correlated with the number of neighbors it has in the specified network, as shown in Figure 4c.

\begin{figure*}
    \centering
    \includegraphics[width=\textwidth]{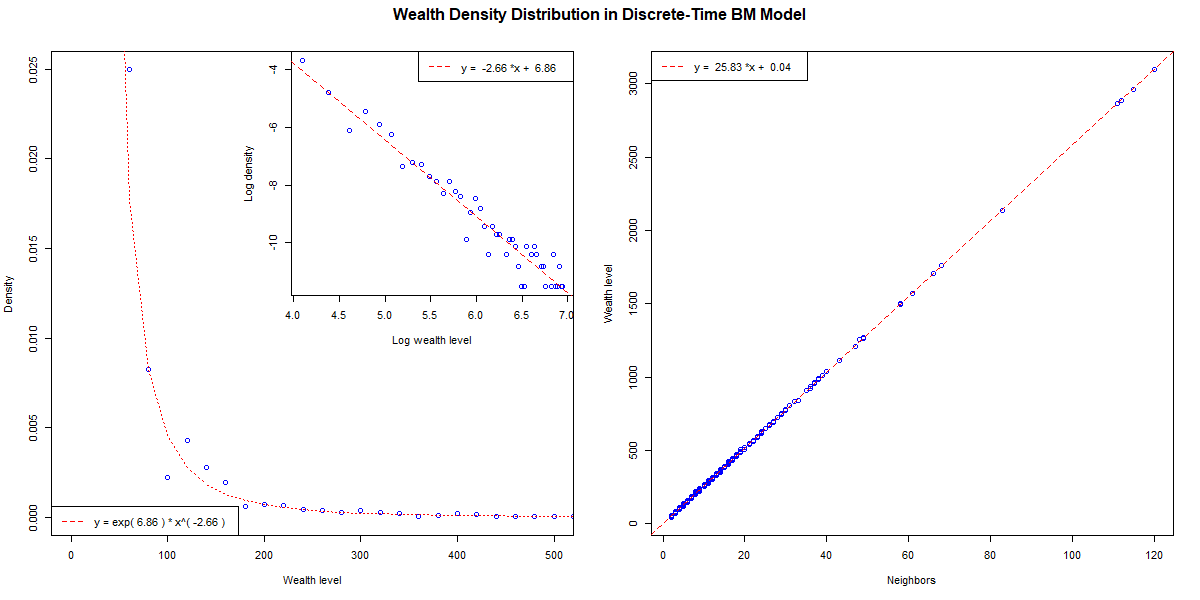}
    \caption[Stationary distribution produced by BM model]{Stationary distribution produced by the discrete-time BM model on a Barab\'{a}si–Albert scale-free network, as described by Di Matteo \textit{et al.} (2003). Simulation was performed with the parameterization \(N = 5000\), \(\langle w \rangle = 100\), \(q_{0} = 0.1\), and \(E \left [A_{i}(t)^{2} \right ] = 1\) over \(10^{5}\) iterations.}
    \label{fig:dimatteo_plot}
\end{figure*}

Scafetta \textit{et al.} (2004) propose another discrete-time variation of Bouchaud and M\'{e}zard's model, motivated by a dissatisfaction with the formulation of wealth transfer via exchange as it appears in their original paper, which sees a constant wealth flux from rich to poor.\nocite{scafetta_trade-investment_2004} This is not necessarily realistic as wealth should only be transferred in exchange if an agent buys an asset for a price different than its value; such a model cannot explain wealth inequality under the assumption of perfect pricing. Thus, Scafetta \textit{et al.} propose a model in which the wealth of agent \(i\) is given by:
\begin{equation}
    w_{i}(t+1) = w_{i}(t) + r_{i}\xi(t) w_{i}(t) + \sum_{j \neq i} w_{i\rightarrow j}(t)
\end{equation}
where \(r_{i} = V\Pi_{i} > 0\) is the "individual investment index," given as the product of the global investment index and the proportion of wealth actually invested by agent \(i\), \(\xi(t)\) is a Gaussian random variable representing return on investment, and \(w_{i \rightarrow j}(t)\) represents the flow of wealth from agent \(j\) to agent \(i\) in period \(t\), which is assumed to be Gaussian with mean \(\mu = fh \frac{w_{i} - w_{j}}{w_{i} + w_{j}} \min\{w_{i}, w_{j}\}\) and standard deviation \(\sigma = h \min\{w_{i}, w_{j}\}\). 

Varying \(f\), \(h\), and \(r\) then allowed the authors to tune the strengths of different system dynamics. If \(h > 0\) and \(f = r = 0\) (the symmetric trade-only model), wealth condensation occurs. If \(f, h > 0\) and \(r = 0\) (the asymmetric trade-only model), a Gamma-like distribution is observed. Finally, if \(f, h, r > 0\) (the asymmetric trade-investment model), a Gamma-like distribution with a power-law tail is observed.

Various other modifications to the BM model have been studied as well. Huang (2004) extends the BM model to negative wealth levels, and Torregrossa and Toscani (2017) prove analytically that a unique steady state with support on the entire real number line exists.\nocite{huang_wealth_2004, torregrossa_wealth_2017} Johnston \textit{et al.} (2005) imposes the additional restriction of conservation of wealth, finding that wealth condensation still occurs for high values of \(\mu\).\nocite{johnston_wealth_2005} Finally Ichinomiya (2012a, b) relaxes Bouchaud and M\'{e}zard's mean field assumption to adiabatic and independent assumptions, drawn from quantum mechanics.\nocite{ichinomiya_bouchaud-mezard_2012, ichinomiya_wealth_2012} The power law-like tail is reproduced and condensation is seen to take place at a higher \(J\) than the mean-field case would indicate, though the Pareto index obtained is smaller than those empirically observed \cite{ichinomiya_power-law_2013}.

\subsection{Other formulations}

While the two principal model classes of the random asset exchange literature are KWE model and the BM model, a variety of less influential formulations also exist. 

A simple model which has nonetheless been significant in the economics literature is the multiplicative stochastic process (MSP), which was studied by the economists Robert Gibrat and D.G. Champernowne \cite{garlaschelli_effects_2008}. It would however be a stretch to say that the MSP class of models is truly a type of random asset \textit{exchange} model as it is primarily characterized by the lack of exchange or any other sort of interaction between agents. Such models essentially represent agents' levels of wealth in terms of independent random walks, but nonetheless are able to capture some essential characteristics of observed distributions. The simplest model in this vein is the pure MSP \(w(t+1) = \lambda(t)w(t)\), where \(\lambda(t)\) is a Gaussian random variable. It is straightforward to show that the distribution of wealth among an ensemble of agents whose wealth evolution is governed by a pure MSP will follow a log-normal distribution, though variations which include additive noise (as in the Kesten process from the biological sciences) and "minimum wage"-style boundary constraints can also reproduce power law tails \cite{souma_physics_2002}. For examples of such models, see Biham \textit{et al.} (1998), Huang and Solomon (2001), Souma and Nirei (2005), and Basu and Mohanty (2008).\nocite{biham_generic_1998, huang_finite_2001, souma_empirical_2005, basu_modeling_2008}

Another class of model which studied in the early days of the RAE literature in particular is the Generalized Lotka-Volterra (GLV) model, which also has its origins in the biological sciences. The original Lotka-Volterra process, studied by Biham \textit{et al.} (1998), is given by:
\begin{equation}
    w_{i}(t+1) = \lambda(t) w_{i}(t) + a\bar{w}(t) - bw_{i}(t) \bar{w}(t)
\end{equation}
where \(\lambda\) is a time-dependent random variable and \(\bar{w}(t)\) is the average wealth in the system.\nocite{biham_generic_1998} The inclusion of the \(\bar{w}(t)\) terms represents a form of indirect interaction between agents: much like in the mean-field approximation of the BM model, instead of including specific interaction terms \(b_{ij}w_{i}(t)w_{j}(t\), all interactions are assumed to be symmetrical: \(b_{ij} = b/N\). 

The generalized form of this model was introduced by Solomon and Richmond (2001, 2002), and follows\nocite{solomon_power_2001, solomon_stable_2002}:
\begin{equation}
    \begin{split}
        \Delta w_{i}(t) &= \left ( \varepsilon_{i}(t) \sigma_{i} + c_{i}(w_{1}, w_{2}, \dots , w_{N}, t) \right ) w_{i}(t) \\ &+ a_{i} \sum_{i} b_{j} w_{j}(t)
    \end{split}
\end{equation}
where \(\epsilon_{i}\) is a stochastic variable such that \(E[\epsilon_{i}] = 0\) and \(E[\epsilon_{i}^{2}] = 1\), \(c_{i}\) represents endogenous and exogenous dynamics in returns, and \(a_{i}\) and \(b_{i}\) represent arbitrary redistributions of wealth among agents. The restrictions on \(\epsilon\) can be made without loss of generality thanks to the \(c_{i}\) term. Under certain assumptions, this model also produced mixed exponential-Pareto distributions. However, this model ultimately faded in popularity due to the difficulty it has accurately representing the left tail of income distributions, as well as the lack of economic justification for some of its terms \cite{repetowicz_dynamics_2005}.

\section{Notable Trends in the Literature}

While the papers discussed above serve as the foundation for the random asset exchange literature, the flexibility of the underlying modeling framework have allowed a vast number of featural variations upon these canonical models to have proliferated. In this section, we provide an overview of the most significant of these trends is provided and summarize a number of key papers in each category. 

\subsection{Non-conservation of wealth}

One of the primary criticisms leveled against the original KWE models is that the assumption of total conservation of wealth, made by analogy with the conservation of energy in ideal gas models, is highly unrealistic. In real economies, wealth is constantly being created and destroyed---not just by means of production and consumption, but even by the constant issuing and repaying of loans. Thus, a number of modifications to the conservative KWE model have attempted to represent this fact. Most such models can be classified into one of two types: models which, like the CPT model, conserve wealth in the mean, and models which tie the global wealth level to a fixed influx rate.

Bisi \textit{et al.} (2009) and Bassetti and Toscani (2010) both consider models of the first type.\nocite{bisi_kinetic_2009, bassetti_explicit_2010} The latter considers the non-conservative exchange rule:
\begin{equation}
    M = 
    \begin{bmatrix}
        \varepsilon_{i} & \varepsilon_{i} \\
        \varepsilon_{j} & \varepsilon_{j}
    \end{bmatrix}
\end{equation}
where \(\varepsilon_{i}\) and \(\varepsilon_{i}\) are i.i.d. and \(E[\varepsilon_{i} + \varepsilon_{j}] = 1\). Bassetti \textit{et al.} (2014) considers a class of similar lotteries and demonstrates they tend to produce inverse-Gamma steady states.\nocite{bassetti_explicit_2014}

Slanina (2004) was the first to consider a non-conservative model of the second type, in which a constant inflow of wealth from outside the system of interacting agents is permitted.\nocite{slanina_inelastically_2004} As in other formulations, the model sees pairs of agents \(i\) and \(j\) chosen at random to engage in a transfer of wealth, defined by the dynamics:
\begin{equation}
    \begin{split}
        M = 
        \begin{bmatrix}
            1 - \lambda + \epsilon & \lambda \\
            \lambda & 1 - \lambda + \epsilon
        \end{bmatrix}
    \end{split}
\end{equation}
where \(\lambda \in [0, 1]\) represents the wealth exchanged between two interacting agents and \(\epsilon > 0\) represents the rate at which exogenous wealth flows into the system. Slanina's model produces a Gamma-like equilibrium distribution with a Pareto tail with an index \(\alpha \sim 1 + \frac{2\lambda}{\epsilon^{2}}\). Coelho \textit{et al.} (2008) extended this model, redefining \(\lambda(w_{i})\) as a piecewise function taking on two different values depending on which side of a pre-specified wealth threshold \(n\bar{w}(t)\) an agent's wealth \(w_{i}(t)\) fell:
\begin{equation}
        M = 
        \begin{bmatrix}
            1 - \lambda(w_{i}(t)) + \epsilon & \lambda(w_{j}(t)) \\
            \lambda(w_{i}(t)) & 1 - \lambda(w_{j}(t)) + \epsilon
        \end{bmatrix}
\end{equation}
This modification reproduced a double power-law regime, a phenomenon observed when comparing the right tail of income from tax data to estimates for the capital gains of a country's very wealthiest individuals.\nocite{coelho_double_2008}

A number of non-conservative models have dynamics which attempt to more directly model the process of money creation through borrowing. For example, Chen \textit{et al.} (2013) consider a random exchange model in which agents who would otherwise reach zero wealth are permitted to borrow money from a central bank, which in turn can issue loans with no interest up to a certain global debt limit.\nocite{chen_money_2013} This process of money creation (issue of loans) and annihilation (paying back of loans) leads to a system in which the money supply grows logarithmically. Schmitt \textit{et al.} (2014) introduces a similar system of money creation and analyzes the non-local effect that issuing credit has on the rest of the system; though the recipient of the loan clearly benefits, the effects of the increase in the money supply quickly propagate and all agents suffer the resultant inflationary effects.\nocite{schmitt_statistical_2014}

Recently, Liu \textit{et al.} (2021) and Klein \textit{et al.} (2021) introduced a generalization of the unaltered YS model which permits growth in the money supply over time, which they call the "Growth, Exchange, and Distribution" (GED) model.\nocite{liu_simulation_2021, klein_mean-field_2021} Each time-step, total wealth \(W(t)\) is increased by a factor of \(1+\mu\), and the wealth influx \(\mu W(t)\) is distributed among agents such that agent \(i\) receives \(w_{i}^{\lambda}/(\sum_{j}w_{j}^{\lambda})\). For subunitary values of \(\lambda\), poorer agents disproportionately benefit from the growth in the money supply and a quasi-stationary distribution exists; otherwise, the system exhibits wealth condensation as in the unaltered YS model. The system dynamics at play here are quite similar to the model of Vallejos \textit{et al.} (2018), in which growth surplus is apportioned according to a more indirect "wealth power" parameter.\nocite{vallejos_agent-based_2018}

\subsection{Networks and preferential attachment}

It is a notable and well-established result that the significance of the introduction of adjacency networks into random asset exchange models depends heavily on the specific model formulation. While, for instance, the specific nature of the network has a decisive effect on the steady state wealth distribution in BM-style models, the opposite tends to be true for KWE-style models. Networks of exchange are an important aspect of real economic systems, and as such there has been a significant effort to study the effect they have on various types of RAE models.

Interestingly, models characterized by unidirectional exchange exhibit greater sensitivity to network structure than bidirectional exchange models. Chatterjee (2009), for example, introduces a toy model in which agents exchange fixed fractions of their wealth on a directed network characterized by a disorder parameter \(p\).\nocite{chatterjee_kinetic_2009} Higher values of \(p\) produced networks where more agents had similar incoming and outgoing connections; the distributions obtained therefrom were more Gamma-like, as opposed to the Boltzmann-like distributions obtained from lower values of \(p\). Martínez-Martínez and López-Ruiz (2013) study a unidirectional model with random exchange fractions, meant to represent payments on a non-complete graph.\nocite{martinez-martinez_directed_2013} This "directed random exchange" (DRE) model thus has the exchange rule:
\begin{equation}
    \begin{split}
        M = \begin{bmatrix} \varepsilon & 0 \\ 1 - \varepsilon & 1 \end{bmatrix} 
    \end{split}
\end{equation}
As with Chatterjee (2009), the choice of adjacency network affects the equilibrium distribution of the DRE model. For the fully-connected case, the equilibrium distribution \(p(x)\) is again exactly Gamma, with shape parameter \(\frac{1}{2}\) \cite{katriel_directed_2015}. Notably, this implies that \(p\) possesses a singularity at 0, explaining why Martínez-Martínez and López-Ruiz observed a condensation-like phenomenon even on fully-connected networks.

S\'{a}nchez \textit{et al.} (2007) investigate a model in which agents populate a one-dimensional lattice. Each agent's wealth grows in a deterministic fashion as a product of a linear "natural growth" term and an exponential "control" term, which retards growth as the difference between an agent's wealth and the average wealth of its neighbors increases.\nocite{sanchez_model_2007} While this system produces a pure power law distribution, it was later demonstrated that, for different values of the system's endogenous parameters or a rearrangement of agents' neighborhoods, either a Boltzmann-Gibbs or a Pareto distribution could be obtained \cite{gonzalez-estevez_pareto_2008, gonzalez-estevez_transition_2009}.

A handful of models have included the additional possibility of agents exchanging connections or positions on a lattice as well as units of wealth. Gusman \textit{et al.} (2005) define an IGAV model on a random network in which the winner of an exchange is rewarded with additional connections on the network, producing a power law regime.\nocite{gusman_wealth_2005} Aydiner \textit{et al.} (2019) examine a CCM-style bidirectional exchange model on a one-dimensional lattice, with the twist that some fraction of agents exchange lattice position each iteration of the simulation.\nocite{aydiner_money_2019} Fernandes and Tempere (2020) likewise consider a variation of the CC model in which agents on a two-dimensional lattice randomly switch positions on the lattice such that the average wealth difference between neighboring nodes is reduced.\nocite{fernandes_effect_2020} This ultimately results in perfect wealth segregation and uniformly higher inequality. 

The dynamics of wealth exchange coupled with extremal dynamics was thoroughly studied in the "conservative exchange market" (CEM) model \cite{pianegonda_wealth_2003, iglesias_wealth_2003, pianegonda_inequalities_2004}. Said model populates a lattice with agents who possess wealth levels in the range \([0, 1]\), and each time step sees the poorest agent's wealth randomly re-randomized at the expense or benefit of its two closest neighbors. The selection rule in this model induces self-organizing behavior such that almost all agents end up with wealth levels above a "poverty line," which proved to be higher in the restricted lattice case than in the fully-connected case. This model has been extended by a number of follow-up papers over the years. Iglesias \textit{et al.} (2010) used this model to compare two different redistribution schemes, and Ghosh \textit{et al.} (2011) considered its mean-field approximation.\nocite{iglesias_how_2010, ghosh_threshold-induced_2011} Chakraborty \textit{et al.} (2012) and Braunstein \textit{et al.} (2013) studied the same dynamics on various other networks.\nocite{chakraborty_conservative_2012, braunstein_study_2013}

\begin{figure*}
    \centering
    \includegraphics[width=\textwidth]{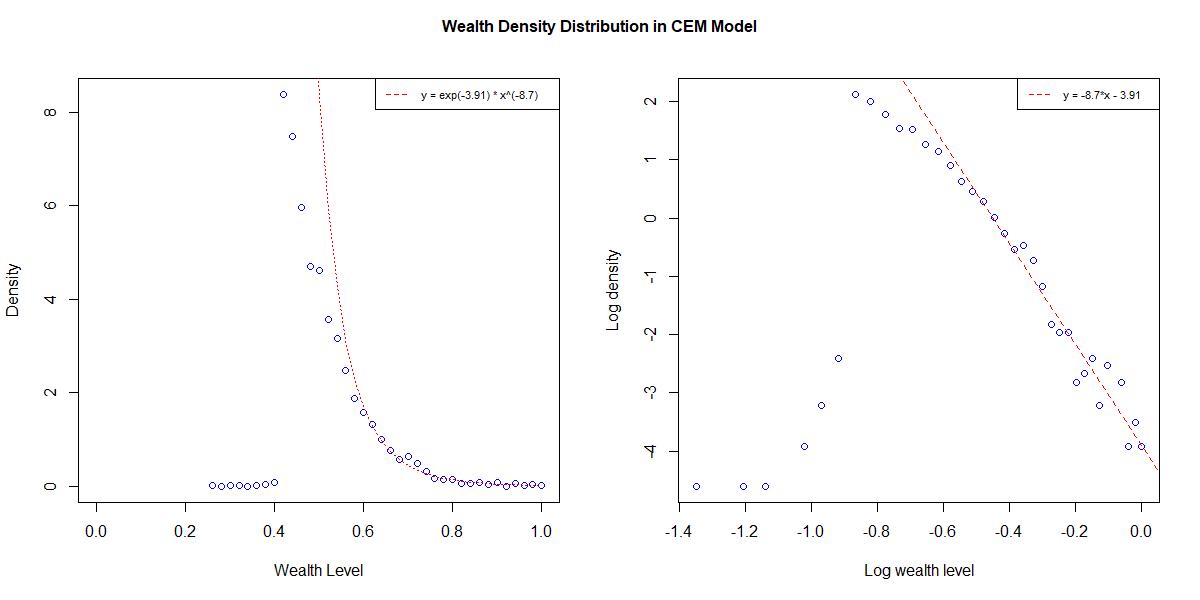}
    \caption[Stationary distribution produced by CEM model]{Stationary distribution produced by the CEM model, with the best log-linear fit of the right tail of the distribution shown. Simulation was performed with \(N = 5000\) over \(10^{7}\) iterations.}
    \label{fig:cemm_plot}
\end{figure*}

A concept closely related to adjacency is that of preferential attachment, which defines the likelihood of two agents interacting as a function of endogenous variables. The variable chosen is usually wealth, representing the fact that, in real economies, both the rich and the poor tend to interact more often with people of similar socioeconomic status to themselves. Because of its non-discrete nature, preferential attachment can allow for somewhat more dynamic interactions than adjacency networks can, permitting agents who become wealthy to access the networks of the rich and not totally disallowing chance rich-poor interactions. In fact, adjacency networks can be viewed as a special case of preferential attachment.

Laguna \textit{et al.} (2005) study the effect of this phenomenon on the IGAV model by imposing the restriction that a given agent is only permitted to interact with another agent if the difference between their two wealth levels is less than a given threshold value \(u\).\nocite{laguna_economic_2005} Large values of \(u\) more or less replicate the IGAV model and small values freeze the system entirely, as one would expect. Intermediate values, however, produced a self-organizing separation within the distribution of wealth, with a gap separating rich agents from poor ones spontaneously arising. This bimodal distribution persisted even for high values of the poor-bias parameter \(f\).

Chakraborty and Manna (2010) introduce a model with simple preferential attachment behavior, such that richer agents engage in exchange more frequently.\nocite{chakraborty_weighted_2010} That is, the probability that agent \(i\) is selected as the first trader is proportional \(w_{i}^{\alpha}\), and the probability that \(j\) is selected as the second trader is proportional \(w_{j}^{\beta}\). The limit as either exponent goes to infinity yields purely extremal mechanics, while \(\alpha = \beta = 0\) is the CCM model. Goswami and Sen (2014) defines a more complicated attachment function, wherein the probability of a given pair of agents \((i,j)\) interacting depends on \(i\)'s total wealth, the difference in wealth between \(i\) and \(j\), and the number of past interactions between agents \(i\) and \(j\).\nocite{goswami_agent_2014} The strength of each factor is modulated by a corresponding exponent, and, when applied to the classical BDY model, the choice of modulation has a significant effect on the Pareto index of the steady-state distribution.

\subsection{Goods and rationality}

Despite their reductive nature, many of the simplifying assumptions discussed above are not uncommon to find in the economics literature as well. Many neoclassical models of simple "exchange economies" study the distribution of endowed and conserved assets absent wealth creation, and comparatively few consider the effect of exchange networks or other kinds of barriers to freely-associating exchange between agents (an important source of imperfect competition and thus market inefficiency). Rather, the main distinction between RAE-style models and those found in the mainstream economics literature lies in the fact that the models preferred in the former typically study ensembles of agents exchanging money directly in a stochastic fashion, while the latter typically study ensembles of rational (i.e. utility-maximizing) agents exchanging goods, with money exchange being an implicit consequence of goods exchange. A number of attempts have been made to partially bridge this difference between these two literatures by introducing goods and rationality into RAE models.

Chakraborti \textit{et al.} (2001) study a model with both a fixed commodity supply \(Q\) and money supply \(M\) distributed among a population of agents.\nocite{chakraborti_self-organising_2001} These agents first seek to ensure their level of goods \(q_{i}\) exceeds some subsistence level \(q_{0}\), and afterwards seek to maximize their money holdings \(m_{i}\); thus, agents with \(q_{i} > q_{0}\) find agents with \(q_{j} < q_{0}\) to sell their excess goods to at a fixed price of 1. Not surprisingly, the steady-state distribution of this system is found to be sensitive to the global quantities \(Q\) and \(N\); if the commodity supply is limited (\(Q/N < q_{0}\)), some fraction of agents will necessarily fall below subsistence level, while if the money supply is limited, agents lack the ability to redistribute the commodity supply in an efficient manner. A similar model with stochastic price fluctuations is considered by Chatterjee and Chakrabarti (2006), in which wealth is taken to be the sum of money and commodity holdings.\nocite{chatterjee_kinetic_2006} In both models, the money distribution exhibits a Pareto tail with index 1 while the commodity distribution is exponential so long as neither \(Q\) nor \(M\) is restricted.

Silver \textit{et al.} (2002) considers a model with a more sophisticated utility function, in which agents possess stochastically time-varying Cobb-Douglas utility functions of the form\nocite{silver_statistical_2002}:
\begin{equation}
    u_{i,t}(a_{i, t}, w_{i, t}) = (a_{i,t})^{f_{i,t}}(w_{i,t} - a_{i,t})^{1 - f_{i,t}}
\end{equation}
where \(a_{i,t}\) represents agent \(i\)'s holdings of the money commodity at time \(t\), \(w_{i,t} - a_{i,t}\) represents agent \(i\)'s holdings of non-money commodities at time \(t\), and \(f_{i,t}\) is a random variable independently and identically distributed across both indices. In an approach highly reminiscent of the derivation of the equilibrium of the canonical Arrow-Debreu model in economics, each agent chooses to re-allocate their wealth between money and non-money commodities in such a way that maximizes \(u_{i,t}\) subject to supply constraints. Simulations of this system produce a wealth distribution well-fit by a Gamma distribution with a shape parameter of 1 and a rate parameter of \(1/\alpha\), where \(\alpha\) represents the global supply of the money commodity.

However, not every exchange model with goods is paired with rational agents. Ausloos and P\c{e}kalski (2007) consider a model with money, goods, and completely stochastic agent behavior.\nocite{ausloos_model_2007} Each time step, one agent decides via coin toss whether to purchase a nonzero number of goods. If so, he randomly selects a fraction of his money to spend and taps another agent to sell to him. If this second agent has enough goods to sell and has a desire to sell (again decided via coin toss), the exchange takes place. This model produces a distribution of wealth which interpolates between two power laws as time progresses, while the distribution of goods follows a static power-law. In general, those agents that are rich in terms of money are poor in terms of goods, and vice versa.

Another interesting line of research has concerned itself with defining traditional macroeconomic ensembles which produce equivalent results to RAE models. For example, Chakrabarti and Chakrabarti (2009) demonstrates that the dynamics of the CCM model can be replicated in a neoclassical framework with rational agents producing differentiated goods and trading in order to maximize time-varying Cobb-Douglas utility functions for goods and money.\nocite{chakrabarti_microeconomics_2009} In this case, the stochastic nature of exchange in the CCM model is represented by random variations in agents' utility functions in the analog model. Tao (2015) derives the entropy-maximizing exponential distribution as the statistical equilibrium of an Arrow-Debreu market system populated by agents with such time-varying utility functions.\nocite{tao_universal_2015} More recently, Quevedo and Quimbay (2020) have extended this formulation to permit agents to save a portion \(s\) of goods possessed, naturally leading to an equivalent non-conservative RAE model.\nocite{quevedo_non-conservative_2020}

\subsection{Strategic behavior}

Another approach to modeling "smarter" agent behavior attempts to integrate game-theoretic or machine learning dynamics into RAE models. The exact nature of this integration can take various forms, including bilateral agreement, strategic heterogeneity, and behavioral evolution, just to name a few.

In Heinsalu and Patriarca's original paper introducing the immediate exchange model, the authors consider the effect of introducing an acceptance criterion---a probabilistic factor defining the odds a given agent will agree to engage in a given transaction as a function of the difference in wealths between the agent and his partner, with both agents needing to agree to a transaction for it to take place \cite{heinsalu_kinetic_2014}. In both the BDY and IE models, the choice of any symmetrical acceptance criterion (whether linear, exponential, etc.) only impacts the time of relaxation to equilibrium, but not the shape equilibrium itself. Asymmetrical decision criteria cause the equilibrium distribution to lose its universal form and to depend instead on the rule chosen. For the CC model, however, introducing even a symmetric criterion causes the equilibrium to lose its Gamma-like shape. 

Sun \textit{et al.} (2008) investigate a KWE model in which each agent can follow one of four strategies, chosen at random before the simulation begins.\nocite{sun_wealth_2008} The exchange rule between two agents depends on their strategy and the strategy of their partner: two of the strategies are passive and tend towards equalizing the wealth of the two agents, while the other two are aggressive and tend towards classical theft-and-fraud exchange. As with Heinsalu and Patriarca (2014), the introduction of heterogeneous trading strategies leads to a steady-state distribution which depends heavily on the model parameters, specifically those defining the rate of success of the aggressive strategies against the passive strategies.

Heterogeneity in strategies is often studied alongside dynamics for updating agents' strategies, representing a rudimentary form of learning. Hu \textit{et al.} (2006, 2007, 2008), for example, consider a model in which agents begin as either cooperators or defectors and play a series of prisoner's dilemma or snowdrift-style games with their neighbors.\nocite{hu_unified_2006, hu_simulating_2007, hu_properties_2008} After each game, an agent identifies the strategy of its richest neighbor and adopts it with some probability defined by his most recent payout, leading on average to more successful strategies propagating throughout the network. In a similar vein, da Silva and de Figueirêdo (2014) investigate an adaptive variation of the CCM model in which each agent \(i\) has a fixed probability \(\gamma_{i}\) of being able to update their savings parameter according to a pre-defined rule each time step.\nocite{da_silva_income_2014} Neñer and Laguna (2021b) study a variation on a poor-biased YS model with non-zero saving propensity, in which a fraction of agents are subjected to a genetic evolutionary algorithm after each Monte Carlo simulation step to update their exchange parameters, which approach the optimal values determined by Neñer and Laguna (2021a).\nocite{nener_wealth_2021, nener_optimal_2021} A BM-style model coupled with game-theoretic dynamics is extensively analytically studied by Degond \textit{et al.} (2014).\nocite{degond_evolution_2014-1}

\subsection{Class division}

As mentioned above, one of the key features of empirical income distributions which RAE models attempt to capture is the bifurcation of the overall distribution into distinct exponential and Pareto ("thermal" and "superthermal," following the terminology of Silva and Yakovenko (2004)\nocite{silva_temporal_2004}) components. While some models attempt to replicate this two-regime behavior while preserving homogeneity of system dynamics (e.g. by distributing a behavioral parameter throughout the population or imposing a specific network structure), a number of authors have instead sought an explanation by means of a bifurcation in system dynamics for agents with large wealth. It is very natural to identify the exponential bulk of the income distribution with labor income and the power law tail with capital gains, seeing as Pareto's original observations came from data for property incomes\nocite{silva_temporal_2004}. In this way, asymmetric system dynamics represent the fact that, in real economies, the rich do indeed have access to economic mechanisms not available to the majority of the population \cite{montroll_introduction_1974}.

Simple models which have this class division "baked in" are easily able to replicate two-regime structures of income. Yarlagadda and Das (2005) and Das and Yarlagadda (2005), for instance, introduce a model in which trading dynamics differ for agents with wealths on either side of a fixed wealth threshold.\nocite{salzano_stochastic_2005, das_analytic_2005} Poorer agents engage in bilateral exchange exactly as in the model of Chakraborti and Chakrabarti (2000), while richer agents engage in exchange---with a different saving parameter---against the system-totality, representing forms of leverage only available to the wealthy. Quevedo and Quimbay (2020) also study a trading model in which a fixed fraction of the population acts as "producers," who employ the remainder of the population as "workers."\nocite{quevedo_non-conservative_2020} Producers trade wealth and pay their associated workers a portion of the exchanged quantity, creating two differently-shaped Gamma distributions for producer and worker income which, when combined, create a clear two-regime distribution. 

Lim and Min (2020) consider the case in which the CCM model is partitioned into two classes by a wealth percentile threshold and a "solidarity effect" among agents below said threshold is introduced.\nocite{lim_analysis_2020} If two agents belong to the same class, then exchange proceeds according to the familiar CCM system dynamics. But if the agents belong to different classes then the lower-class agent gathers "partners" equal to some fraction of the size of the class, and wins a fraction of the upper-class agent's wealth with a probability equal to the percent wealth his coalition possesses in the exchange. This solidarity factor turns out to be crucial for the generation of a realistic wealth distribution, as without it the middle income stratum collapses and one obtains a bimodal distribution, as with Laguna \textit{et al.} (2005).

Imposing a fixed boundary differentiating the upper class from the lower is not necessarily the best approach here, however, as analysis has shown that the "superthermal" component of the income distribution is highly volatile, fluctuating in size with the stochastic movements of financial markets \cite{silva_temporal_2004}. A number of models consequently attempt to capture this out-of-equilibrium aspect of the distribution's right tail by setting class boundaries dynamically. Russo (2014) investigates a model without exchange in which a new wealth percentile threshold defining the size of the upper class is chosen from the uniform distribution at each time step.\nocite{russo_stochastic_2014} Agents above that threshold then see their wealth augmented by a multiplicative stochastic process, while agents below it have their wealth augmented by an additive stochastic process. A different approach is forwarded by Smerlak (2016), who constructs a Markov process defining transition probabilities between a finite number of stratified classes.\nocite{smerlak_thermodynamics_2016} Agents in higher classes derive proportionally greater amounts of income from a multiplicative process subject to shocks, and consequently exhibit much greater fluctuations in wealth compared to the majority of agents, who persist at low levels of wealth indefinitely.

Finally, we wish to highlight here the unique and striking "social architecture" (SA) model of Wright (2005), which sees agents spontaneously self-organize into three distinct classes.\nocite{wright_social_2005} Wright defines an ensemble with three types of agents---employers, employees, and the unemployed---and in each iteration, an agent \(i\) is randomly chosen to be "active." The activities agent \(i\) engages in depends on its status: if \(i\) is an employer, it pays as many of its employees as it can afford; if \(i\) is an employee, it receives a wage and spends it on consumption goods produced by an employer; and if agent \(i\) is unemployed, a random (wealthy) agent is chosen to hire \(i\), assuming their level of wealth is sufficient to pay \(i\)'s wages. Although the initial conditions of the simulation posited complete equality of agents (all agents began with equal wealth and no employer or employees), the population quickly restructured itself into a three-class regime with a distribution of wealth characterized by an exponential bulk and a Pareto tail. The exact nature of this distribution becomes clear when disaggregated for class: the wealth of "employee" agents was completely governed by an exponential distribution, while that of "employer" agents was well-fit by a power law. This result is consonant the argument forwarded by Montroll and Shlesinger (1982) and in contradistinction to explanations of the two-regime distribution which rely on endogenous differences between agents. Unfortunately, Wright's model has seen few direct extensions, though a similar self-organizing model was studied in Lavička \textit{et al.} (2010).\nocite{lavicka_employment_2010}

\begin{figure*}
    \centering
    \includegraphics[width=\textwidth,height=\textheight]{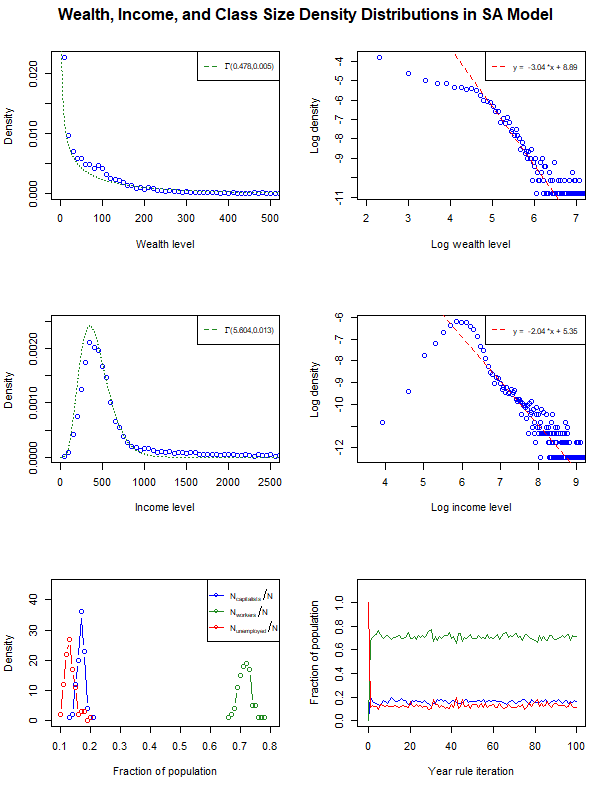}
    \caption[Stationary distributions produced by SA model]{Stationary distributions of wealth, income, and class size produced by the SA model of Wright (2005). Simulation was performed with the parameterization \(N = 5000\), \(\langle w \rangle = 100\), \([w_{a}, w_{b}] = [10, 90]\),  over \(10^{2}\) "year rule" iterations (\(6 \cdot 10^{6}\) time-steps).}
    \label{fig:wright_plot}
\end{figure*}

\subsection{Taxation and redistribution}

A good deal of attention has also been dedicated to the potential usefulness of RAE models in the analysis of the efficiency of redistributive mechanisms. Early studies such as Guala (2008) and Toscani (2009) considered the effect of a simple "income tax," in which a fixed fraction is withdrawn from each exchange by an external body and uniformly redistributed, on mean-conservative KWE models, which was found to not alter the exponential nature of the steady state distribution.\nocite{guala_taxes_2008, toscani_wealth_2009} Diniz and Mendes (2012) extend this result to multiple different taxation rules on a CC model, representing both income taxes (taxes on transaction amounts) and wealth taxes (taxes on wealth level).\nocite{diniz_effects_2012} Bouleau and Chorro (2017) contrast the effect of income and wealth taxes on YS-like models, demonstrating analytically that the income taxes alone are not sufficient to prevent condensation.\nocite{bouleau_impact_2017} Similarly, Burda \textit{et al.} (2019) investigate the dynamics of a BM-style model with the parameterization \(J<0\), which would normally cause the system to condense, paired with a redistributive mechanism.\nocite{burda_dynamics_2019} A sufficiently strong mechanism succeeded in preventing condensation and recovering a heavy-tailed wealth distribution, with a multimodal critical phase also being observed. A number of non-standard redistribution rules in a YS model were examined by Lima \textit{et al.} (2022).\nocite{lima_nonlinear_2022}

Recently, however, interest within the literature has grown around the problem of identifying optimal tax rates in models, often borrowing techniques from control theory to do so. Bouchaud (2015) extends the BM model to permit a wealth tax capable of reallocating wealth between the private and public sectors, with different growth rate parameters.\nocite{bouchaud_growth-optimal_2015} By maximizing expected economic growth, an optimal tax rate in the interval \((0, 1)\) is obtained for growth rate differences within an intermediate range. D\"{u}ring \textit{et al.} (2018) develop a finite-horizon model predictive control mechanism for the CPT model to derive the feasible tax regime which minimizes a cost function representing some metric of inequality, and consider various objective functions and redistribution schemes.\nocite{during_kinetic_2018} Zhou and Lai (2022) investigate a novel model of individual wealth growth and formulate both an additive and a multiplicative control mechanism to modulate the excessive growth of the right tail of the wealth distribution of the ensemble.\nocite{zhou_kinetic_2022} Lastly, Wang \textit{et al.} (2022) pairs a CPT model with an evolutionary description of agents' decision-making competence---which feeds back into their saving propensities---and a model predictive control mechanism to reduce inequality.\nocite{wang_optimal_2022}

\begin{table*}[t]
\scriptsize
    \centering
\begin{tabular}{ |p{1.5cm}||p{3cm}|p{3cm}|p{3cm}|p{3cm}| }
 \hline
 \multicolumn{5}{|c|}{\textbf{Notable Papers}} \\
 \hline
 \textit{Model \newline Feature} & \textit{Theft \& Fraud} & \textit{Yard Sale} & \textit{Bouchaud-M\'{e}zard} & \textit{Other} \\
 \hline
 Canonical & \text{Angle (1986)} \newline \text{Bennati (1988)} \newline \text{Ispolatov \textit{et al.} (1998)} \newline \text{Dr\u{a}gulescu \&} \newline \text{Yakovenko (2000)} \newline \text{Chakraborti \&} \newline \text{Chakrabarti (2000)} \newline \text{Chatterjee \textit{et al.} (2003)} & \text{Hayes (2002)} \newline \text{Iglesias \textit{et al.} (2004)} \newline \text{Caon \textit{et al.} (2007)} \newline \text{Moukarzel \textit{et al.} (2007)} & \text{Bouchaud \& M\'{e}zard} \newline \text{(2000)}  & \text{Biham \textit{et al.} (1998)} \newline \text{Solomon \& Richmond} \newline \text{(2001, 2002)} \\
 \hline
 Non-cons. & \text{Slanina (2004)} \newline \text{Cordier \textit{et al.} (2005)} \newline \text{Coelho \textit{et al.} (2008)} \newline \text{Bisi \textit{et al.} (2009)} \newline \text{Bassetti \& Toscani} \newline{(2010)} \newline \text{Chen \textit{et al.} (2013)} \newline Schmitt \textit{et al.} (2014) & \text{Liu \textit{et al.} (2021)} \newline \text{Klein \textit{et al.} (2021)} & \text{ } & \text{Heinsalu \& Patriarca} \newline \text{(2014)}\\
 \hline
 Networks & \text{Chatterjee (2009)} \newline \text{Mart\'{i}nez-Mart\'{i}nez \&} \newline \text{L\'{o}pez-Ru\'{i}z (2013)} \newline \text{Aydiner \textit{et al.} (2019)} \newline \text{Fern\&es \& Tempere} \newline \text{(2020)} & \text{Gusman \textit{et al.} (2005)} \newline \text{Laguna \textit{et al.} (2005)} \newline \text{Guajardo \&} \newline \text{Moukarzel (2012)} & \text{Souma \textit{et al.} (2001)} \newline \text{Di Matteo \textit{et al.} (2003)} \newline \text{Scafetta \textit{et al.} (2004)} \newline \text{Garlaschelli \&} \newline \text{Loffredo (2004)} \newline \text{Ma \textit{et al.} (2013)} & \text{Pianegonda \textit{et al.}}\newline \text{(2003)} \newline \text{S\'{a}nchez \textit{et al.} (2007)}\\
 \hline
 Goods & \text{Chakraborti \textit{et al.}} \newline \text{(2001)} \newline \text{Chatterjee \& } \newline \text{Chakrabarti (2006)} & \text{ } & \text{ } & \text{Ausloos \& P\c{e}kalski} \newline \text{(2007)}\\
 \hline
 Rationality & \text{Chakrabarti \& } \newline \text{Chakrabarti (2009)} \newline \text{Tao (2015)} \newline \text{Quevedo \& Quimbay} \newline \text{(2020)} & \text{ } & \text{ } & \text{Silver \textit{et al.} (2002)}\\
 \hline
 Strategies & \text{Sun \textit{et al.} (2008)} \newline \text{da Silva \&} \newline \text{de Figueirêdo (2014)} & \text{Neñer \& Laguna} \newline \text{(2021b)} & \text{Degond \textit{et al.} (2014)} & \text{Hu \textit{et al.} (2006)}\\
 \hline
 Class div. & \text{Yarlagadda \& Das} \newline \text{(2005)} \newline \text{Lim \& Min (2020)} & \text{ } & \text{ } & \text{Wright (2005)} \newline \text{Lavička \textit{et al.} (2010)} \newline \text{Russo (2014)} \newline \text{Smerlak (2016)}\\
 \hline
 Redist. & \text{Guala (2008)} \newline \text{Toscani (2009)} \newline \text{Diniz \& Mendes (2012)} & \text{Boghosian (2014a)} \newline \text{Boghosian \textit{et al.} (2017)} \newline \text{Bouleau \& Chorro} \newline \text{(2017)} \newline \text{D\"{u}ring \textit{et al.} (2018)} \newline \text{Lima \textit{et al.} (2022)} \newline \text{Wang \textit{et al.} (2022)} \newline \text{Li \textit{et al.} (2019)} & \text{Bouchaud (2015)} \newline \text{Burda \textit{et al.} (2019)} & \text{Zhou \& Lai (2022)}\\
 \hline
\end{tabular}
    \caption[Notable papers in the random asset exchange literature]{Notable papers in the random asset exchange literature, disaggregated by formulation type and prominent features.}
    \label{tab:big_table}
\end{table*}[b]

\subsection{Miscellanea}

Though the above list enumerates the most widely-studied modifications to RAE models, it should by no means be considered exhaustive. The flexibility of the random asset exchange framework makes it easy to introduce new system dynamics and isolate the effects of a given modification. Just to name a few examples, Pareschi and Toscani (2014) investigate the effect of variable agent knowledge on the CPT model, obtaining the intriguing result that the most knowledgeable agents tend not to be the richest ones; Trigaux (2005) examines the effect of introducing altruistic behavior to a subpopulation and finds a very strong equalizing effect when combined with redistribution; Coelho \textit{et al.} (2005) and Patrício and Araújo (2021) model the propagation of wealth on a generational network to study the stratifying effect of inheritance; and Dimarco \textit{et al.} (2020) use a class-based framework to characterize the effect of pandemics on wealth inequality.\nocite{pareschi_wealth_2014, trigaux_wealth_2005, coelho_family-network_2005, patricio_inheritances_2021, dimarco_wealth_2020}.

The RAE literature has also given rise to a number of wholly new analytical techniques. Ballante \textit{et al.} (2020) demonstrate that fitting the distribution of saving propensities to real-time economic data in a generalized CCM model via statistical sampling may be useful as a leading indicator of economic stressors which have the potential to increase inequality.\nocite{ballante_economic_2020} Luquini \textit{et al.} (2020) establish a formal equivalence between KWE models and population-based random search algorithms in computer science, and speculate that said formulation could ultimately be used as a benchmark model in cybernetics.\nocite{luquini_fusing_2020} Finally, dos Santos \textit{et al.} (2022) propose a computational technique by which the crossover point between the exponential and Pareto regimes can be identified within data sets of real income distributions, aiding in the empirical study of economic inequality.\nocite{dos_santos_optimal_2022}

\section{Discussion}

From the ambition and breadth of recent publications such as those mentioned above, it is clear that random asset exchange modeling is being increasingly recognized as a highly versatile tool which has the potential to find wide application even beyond its original use as a descriptive econophysical model. It is also clear that, in seeking to explain the characteristic features of wealth and income distributions, such models have highlighted the existence of a number of more fundamental economic phenomena underlying those features, such as the inherently diffusive nature of exchange economies and the emergence of apparent power laws from overlapping exponential functions.

Furthermore, it has also become apparent in the course of this investigation that the random asset exchange modeling literature has a number of bigger-picture implications. Namely, all of the models discussed indicate that a large proportion of observed economic inequality is the result of luck and the inherently diffusive (entropy-increasing) nature of exchange itself. While some authors have taken this to mean that the "natural," entropy-maximizing level of inequality is by definition fair, such a conclusion is far too strong and veers into the territory of naturalistic fallacies. Instead, the conclusion one ought to draw from this cardinal result of the random asset exchange literature depends on one's own (subjective) beliefs concerning the "ideal" level of inequality---however that is determined---as compared to the level currently prevailing. For proponents of relatively unrestrained capitalism, who have argued that inequality plays an important role in the economy by encouraging people to work harder in the hopes of achieving better economic outcomes, the implications of said result are quite positive: statistics seem to naturally guarantee such inequality without the help of market-distorting conditions such as the formation of monopolies or the institutionalization of economic thievery! On the other hand, for those policy makers who aim to reduce the degree of inequality in modern, developed economies, the corresponding implication may be somewhat more dismal. For them, the main implication of these models is that altering government policies to make market economies operate more "fairly" by, for example, introducing progressive taxation can only do so much. At the end of the day, large scale regimes of wealth redistribution, such as wealth taxes, may be necessary in order to reduce inequality below the level that is endogenous to exchange-based systems.

However, econophysical models are not without their own problems. Most are still incapable of replicating all of the characteristic features of wealth and income distributions. For wealth distributions, as has been discussed, these include non-negligible segments of the population with non-positive wealth and possibly a power law right tail; for income distributions, these include an exponential or log-normal bulk, and an at least apparent power law tail with exponent between -2 and -3. 

More importantly, while all of the models discussed above serve as excellent demonstrations of the role random chance plays in generating the inequalities observed in market economies, the literature has not yet been able to provide an adequate \textit{explanation} for the emergence of these distributional features which it posits to be universal. That is to say, it has not yet been able to identify and describe the concrete system dynamics, common to all market economies, which generate the characteristic features of inequality.

Models which impose specific distributions on an endogenous parameter throughout the population (thriftiness, size of social network, etc.) clearly have the capability of producing nearly any desired distribution, but such results have far less explanatory power seeing as they merely defer the question. If one observes a given distribution of wealth because there exists an underlying distribution of a certain behavioral parameter, why is this parameter distributed the way it is throughout the population? One ultimately returns to Pareto's unsatisfying explanation for his own law---that economic inequality is purely the result of intrinsic differences between individuals---and finds oneself no closer to actually understanding the crux of the issue. 

More promise is shown by models which offer, as Reddy (2020) calls it, a "processual" account of inequality \cite{reddy_what_2020}. Such models introduce the elements of production and class relationships into the mix as fundamental processes of economic systems. This approach reflects concrete asymmetries in the economy, reduces the degree of reductionism present within the models, and permits the identification of different sections of wealth and income distributions with different social positions. Unfortunately, only a handful of studies in this direction have been performed, with Wright (2005) and Lavička \textit{et al.} (2010) remaining the two most notable examples.\nocite{wright_social_2005, lavicka_employment_2010}

Another significant problem pertains to the relationship between the distribution of wealth and the distribution of income, the nature of which the literature has not consistently grasped. Wealth and income are two linked but quite distinct quantities. Wealth can take a wide variety of forms---money, consumption goods, real estate, debts, and even information or skills can all be considered forms of wealth. Income, on the other hand, typically refers to the amount of "wealth," however it is enumerated, received by an individual in a given time period, prior to expenses. This quantity is of course added to one's existing wealth, but the straightforward relation of income as the time-derivative of wealth only holds under the simplifying assumption that wealth is not subject to any endogenous changes: that is, no articles of wealth are consumed, fluctuate in value, are traded for articles of differing value, etc.

For the most part, random asset exchange models are concerned with the distribution of an undifferentiated, non-consumable, exchangeable asset---usually a stand-in for money---throughout an ensemble of agents. Thus, the distributions of said asset throughout the population are best interpreted as wealth distributions, and it is inapt to compare them to empirical distributions of income. In fact, Xu \textit{et al.} (2010) note that reconstructing time-series of agents' income within canonical KWE models actually produces income distributions which are Gaussian, as opposed to exponential, directly contrary to the available data.\nocite{xu_income_2010} Once again, a greater focus on developing models in a more processual vein, which explicitly link income to salaries and wages paid out by firms to employees, could be immensely useful in clearing up this confusion.

Further progress towards an econophysical explanation for inequality is sorely needed. Per Horowitz \textit{et al.} once again, 61\% of American adults believe that there exists too much inequality in the United States today. \cite{pew_inequality_2020} Of that number, 81\% believe that this problem will require either major policy interventions or a complete restructuring of the economy to address. There exists a clear political will, at least in the U.S., to reduce the degree of inequality that has been allowed to develop over the past few decades. But despite that fact, the same Pew survey demonstrated that there exists no consensus on what the major contributors to economic inequality in the U.S. even are. While there are a number of potential explicators commonly cited---such as industrial outsourcing, the country's tax structure, and intrinsic differences between individuals---no single one is viewed by a majority of the population as a decisive factor. Needless to say, determining what policies would be required to create a fairer society and economy necessitates a clearer understanding of the principal processes responsible for generating inequality in the first place, and much work remains to be done before such a satisfactory understanding is reached.

\bibliography{bibliography}

\end{document}